\documentclass[aps,twocolumn,showpacs,floatfix]{revtex4}

\usepackage{epsfig}

\newcommand{\beq}{\begin{equation}}
\newcommand{\eeq}{\end{equation}}
\newcommand{\bea}{\begin{eqnarray}}
\newcommand{\eea}{\end{eqnarray}}
\newcommand{\hf} {\frac{1}{2}}
\newcommand{\nn}{\nonumber\\}

\newcommand\eqn[1]{Eq.\,(\ref{#1})}
\newcommand\eqns[2]{Eqs.\,(\ref{#1}) and~(\ref{#2})}

\newcommand\fig[1]{Fig.\,{\ref{#1}}}
\newcommand\sect[1]{Sect.\,{\ref{#1}}}
\newcommand\sects[2]{Sects.\,\ref{#1} and~\ref{#2}}

\newcommand\tab[1]{Table~\ref{#1}}

\def\t{\tilde}
\def\tV{\tilde V}

\begin{document}

\title{Critical exponents of the $O(N)$ model in the infrared limit
from functional renormalization}

\author{S. Nagy}
\affiliation{Department of Theoretical Physics, University of Debrecen,
P.O. Box 5, H-4010 Debrecen, Hungary}
\affiliation{MTA-DE Particle Physics Research Group, P.O.Box 51,
H-4001 Debrecen, Hungary}

\begin{abstract} 
We determined the critical exponent $\nu$ of the scalar $O(N)$ model with
a strategy based on the definition of the correlation length in the infrared limit.
The functional renormalization group treatment of the model shows that
there is an infrared fixed point in the broken phase.
The appearing degeneracy induces a dynamical length scale there, which can be considered as
the correlation length. It is shown that the IR scaling behavior can account either
for the Ising type phase transition in the 3-dimensional $O(N)$ model, or for
the Kosterlitz-Thouless type scaling of the 2-dimensional $O(2)$ model.
\end{abstract}
\pacs{11.10.Gh, 05.10.Cc, 64.60.Ak}

\maketitle

\section{Introduction}

The renormalization group (RG) method is a powerful tool to take into account
all the quantum fluctuations systematically \cite{Wetterich,effRG}. The RG flow
equations constitute a bridge between the high energy, microscopic, ultra-violet
(UV) physics and the low-energy, infrared (IR) one. One usually starts with a UV
potential, which describes the small distance interactions among the
elementary excitations and, by taking into account the quantum fluctuations in
a consecutive manner, one finally arrives at the effective theory which contains
the complete effects of the quantum modes.

The RG flow equations are highly non-linear and, due to the complexity of
initial conditions, it is difficult to map out the whole phase space, therefore
one should use some approximations. In the vicinity of the fixed points of the
RG equations one can linearize the flow equations which can make the physical problem
analytically treatable there. The 3-dimensional (3d)
$O(N)$ symmetric scalar model contains a Gaussian and a Wilson-Fisher (WF)
fixed point. Novel approaches and new improvements of the RG method are usually
tested by calculating the scaling exponents in the vicinity of its non-trivial WF fixed
point. The exponents can be calculated numerically by field expansion of
the potential \cite{Tetradis_1994,Liao,exps_field}, or e.g. by $\epsilon$-expansion in
$4-\epsilon$ dimensions \cite{exps_eps}, furthermore one can investigate the convergence
the value of the exponents in the derivative expansion \cite{exps_der}.
In the framework of the RG method more precise results can be
obtained without expanding the potential \cite{exps_shoot,Pangon}.
The exponents can also be calculated by improved Hamiltonian for the 3d
Ising model \cite{exps_ham}.

Our goal is to show that there is an IR fixed point in the d-dimensional $O(N)$ models
inducing an IR scaling regime, which enables one to determine the correlation length
and the critical exponents there. Since the IR fixed point is attractive,
the scaling of the correlation length $\xi$ cannot be determined with the usual technique
according to the linearization around the saddle-point-type WF fixed point.
We use an non-conventional strategy in order to define $\xi$ which is based on
the IR behavior of the RG flows in the broken phase \cite{Nagy_2010,Braun}.
It was shown \cite{Wetterich_kink,Alexandre,Boyanovsky} that at a certain momentum scale the RG
evolution stops that defines a maximal length scale which can characterize the
appearing global condensate there. This scale can be identified by the correlation length $\xi$.
It diverges as a power law-function which can provide us the critical exponent $\nu$ of $\xi$
in the IR limit. This method enabled one to determine the value of $\nu$ of the 3d scalar
$O(1)$ model, and to recover the scaling of the Kosterlitz-Thouless
(KT) -type phase transition in the sine-Gordon (SG) model, too \cite{Nagy_2010}.

We use the functional RG treatment for the d-di\-mensional $O(N)$ model
for its effective average action in order to
get the deep IR scaling behavior of the effective potential.
We show that the exponent $\nu$ based on the power law scaling of the
correlation length in the IR limit for the 3d $O(N)$ model and for the $O(1)$ model
with continuous dimension coincides with the one, which can be obtained around the WF fixed
point by the conventional method based
on finding negative reciprocal of the eigenvalue corresponding to the
single relevant operator. We also show that the other critical exponents are not
necessarily equal in these regions, e.g. the exponent $\eta$ calculated in the IR scaling regime
is much larger than its value determined in the vicinity of the WF fixed point.
Furthermore, we investigated the 2d $O(2)$ model, where instead of
the WF fixed point a critical slowing down of the evolution appears.
We obtained that the IR scaling reproduces properly the KT-type scaling of the model
in a very simple way. It is important to note that our treatment can determine
the value of the exponent $\nu$ for the KT-type scaling directly, which is in principle
not possible in the conventional way.

The paper is organized as follows. In \sect{sec:ren}
we give the RG evolution equation of the scalar $O(N)$ model.
In \sect{sec:phase} the typical phase structure of the $O(N)$ model is presented and
the appearing fixed points are discussed.
The results for the exponent $\nu$ for the 3d $O(N)$ model and for the $O(1)$ model
with continuous dimension are shown in \sects{sec:ond}{sec:dnu}.
We recover the essential scaling of the correlation length in the IR limit
for the 2d $O(2)$ model in \sect{sec:o2d2}, and finally, in \sect{sec:sum}
the conclusions are drawn up.

\section{Renormalization}\label{sec:ren}

The successive elimination of the quantum fluctuations is performed by means
of the Wetterich RG equation for the effective action \cite{Wetterich}
\beq\label{WRG}
k\partial_k\Gamma_k=\hf\mbox{Tr}\frac{k\partial_k R_k}{R_k+\Gamma_k''}
\eeq
where the prime denotes the differentiation with respect to the field variable $\phi$,
and the trace Tr denotes the integration over all momenta and the summation over the
internal indices.
\eqn{WRG} has been solved over the functional subspace defined by the ansatz
\beq\label{eaans}
\Gamma_k = \int_x\left[\frac{Z}2 \partial_\mu\phi^a\partial^\mu\phi_a + V\right],
\eeq
with the wavefunction renormalization $Z$ and the potential $V$, which are functions of
the invariant $\rho=\phi^a\phi_a/2$. The potential has the form
\beq\label{dimpot}
V = \sum_{n=2}^M\frac{g_n}{n!}(\rho-g_1)^n,
\eeq
with $M$ the degree of the Taylor expansion and the dimensionful
couplings $g_n$, $n\ge 1$. One can introduce the dimensionless couplings
according to $g_1=k^{[g_1]}\kappa$ and $g_n=k^{[g_n]}\lambda_n$, with
$[g_1]=d-2$ and $[g_n]=d+n(1-d/2)$ for $n\ge 2$. For shorthand we use $\lambda_2=\lambda$.
The further dimensionless quantities are denoted by $\sim$, e.g.
$V=k^d \tilde V$, thus
\beq\label{pot}
\tilde V = \sum_{n=2}^M\frac{\lambda_n}{n!}(\rho-\kappa)^n.
\eeq
The introduction of the coupling $\kappa$ serves a better convergence in the broken
phase. The IR regularization is chosen to be the Litim's regulator \cite{Litim_opt,Litim}
\beq\label{litreg}
R_k = (k^2-q^2)\theta(k^2-q^2),
\eeq
which gives fast convergence during the evolution, and provides simple analytic forms
for the flow equations. The evolution equation for the potential can be derived
from \eqn{WRG} \cite{Tetradis_1994}, which reads as \cite{Litim_opt}
\bea\label{potev}
k\partial_k\tV &=& -d \tV+(d-2+\eta)\t\rho\tV'+\frac{4v_d}{d}\left(1-\frac{\eta}{d+2}\right)\nn
&&\times\left(\frac1{1+\tV'+2\t\rho\tV''}+\frac{N-1}{1+\t V'}\right),
\eea
with the Litim's regulator in \eqn{litreg}, where $'=\delta/\delta\rho$
and $v_d=1/\Gamma(d/2)2^{d+1}\pi^{d/2}$, with the Gamma function $\Gamma(d/2)$.
In \eqn{potev} we introduced the anomalous dimension $\eta$ which is defined as
$\eta=-d\log Z/d\log k$ and can be calculated by means of the couplings as
\beq\label{etak}
\eta = \frac{16 v_d}{d}\frac{\kappa\lambda^2}{1+2\kappa\lambda}
\eeq
if we consider the Litim's regulator. The inclusion of $\eta$ in the RG equation accounts for
the evolution of the wavefunction renormalization. We note that a more
precise treatment can be obtained if one Taylor expand $Z$ as the potential $V$ in \eqn{pot},
and considers its evolution equation which has a similar but more involved structure as
the potential has in \eqn{potev}. We also note that the inclusion of the full
momentum dependence of the wavefunction renormalization \cite{full_mom}
would make the behavior of the anomalous dimension regular, which could show that the
obtained scaling of $\eta$ is the consequence of the approximations used here.

From the functional RG equation in \eqn{potev} one can deduce
evolution equations for the couplings $\kappa$, $\lambda$ and $\lambda_n$, $n\ge 2$, e.g.
in $d=3$ the flow equations are
\bea\label{klflow}
k\partial_k \kappa &=& -\kappa+\frac1{2\pi^2(1+2\kappa\lambda)^2}\nn
k\partial_k \lambda &=& -\lambda+\frac{3\lambda^2}{\pi^2(1+2\kappa\lambda)^3}
\eea
for the first two couplings if we set $\eta=0$ and $\lambda_n=0$ for $n> 2$.
The scale $k$ covers the momentum interval from the UV cutoff $\Lambda$ to zero.
In numerical calculations we typically set $\Lambda=1$.
The dependence of the number of couplings should always be considered
\cite{Tetradis_1994,Liao,Bervillier}. We also investigated the $M$ dependence of
our numerical results, and we obtained that the choice $M=8$ gives stable
results for the exponents in the IR limit if we Taylor expand the potential.

\section{The phase space, fixed points}\label{sec:phase}

\begin{figure}
\begin{center} 
\epsfig{file=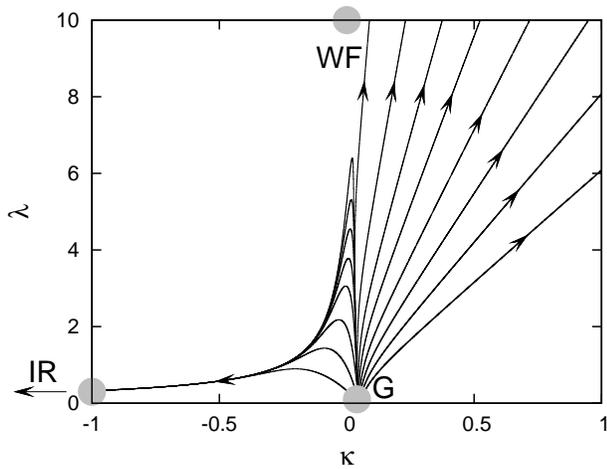,width=6cm,angle=-90}
\caption{\label{fig:phase} The phase space of the 3d $O(N)$ model.
The flows belonging to the symmetric (broken) phase tend to right (left),
respectively.
} 
\end{center}
\end{figure}

The $O(N)$ model in $d=3$ has two phases. The typical phase structure is depicted in
\fig{fig:phase} for the couplings in \eqn{klflow}.
There are two fixed points in the model, which can be easily identified from the RG
equations in \eqn{klflow} if one solves the static equations.
The Gaussian fixed point is situated near the origin
at $\kappa_G^*=1/2\pi^2$ and $\lambda_G^*=0$. The linearization of the flow equations
around the Gaussian fixed point gives a matrix with positive eigenvalues ($s_{G1}=1$ and
$s_{G2}=1$) for the couplings showing that this fixed point is repulsive,
or UV attractive. The WF fixed point can be found at the values
of the couplings $\kappa^*_{WF}=2/9\pi^2$ and $\lambda^*_{WF}=9\pi^2/8$, with eigenvalues
$s_{WF1}=1/3$ and $s_{WF2}=-2$. The opposite signs of the eigenvalues refer to the hyperbolic, or
saddle point nature of the fixed point with attractive and repulsive directions in
the phase space.

We usually identify the critical exponent $\nu$ of the correlation length $\xi$
in the vicinity of the WF fixed point by taking the negative reciprocal of the single negative
eigenvalue, which gives $\eta_{WF}=1/2$ in this case. For many couplings we should find the
fixed points by solving the static RG equations numerically.
These equations are non-linear, and it is very difficult to find them in an $M$ dimensional
phase space, especially if one does not know in which region of the space
space they can be found. Conventionally the fixed points are found by  e.g. shooting method
where the non-expanded potential is considered. However this technique is inappropriate
for models where only a critical slowing down appears in the RG flow, since no fixed points
can be found. This situation takes place e.g. in the 2d $O(2)$ model.

It is worthwhile to mention
that the flows tend to a single curve beyond the WF fixed point. In the broken phase
one obtains a (super)universal effective potential. The phase structure suggests in \fig{fig:phase}
that a further
fixed point may exist in the IR limit. The IR effective potential in the broken
phase can be characterized by an upside down parabola \cite{Wetterich_kink,Alexandre}
which implies that $\kappa_{IR}^*\to-\infty$ and $\lambda_{IR}^*=0$,
therefore the IR limit should be found there. Mathematically it seems that there are no further
fixed points of the RG equation in \eqn{klflow}, but the values for $\kappa_{IR}^*$ and
$\lambda_{IR}^*$ could satisfy the static equations, although they give expressions like
$\infty-\infty$ and $0/0$.
However if one rescales the couplings as in \cite{Tetradis,Nagy_ZSG} then the attractive IR
fixed point can be uncovered. If one repeats those calculation for the couplings in \eqn{klflow}
one finds the following pair of evolution equations
\bea
\partial_\tau \omega &=& 2\omega(1-\omega)-\frac{\ell\omega}{\pi^2}(3-4\omega)\nn
\partial_\tau \ell &=& \ell(5\omega-6)+\frac{9\ell^2}{\pi^2}(1-\omega),
\eea
where $\omega=1+2\kappa\lambda$, $\ell=\lambda/\omega^3$ and $\partial_\tau=\omega k\partial_k$.
The static equations now have the Gaussian ($\ell_G^*=0$, $\omega_G^*=1$),
the WF ($\ell_{WF}^*=\pi^2/3$, $\omega_{WF}^*=3/2$) and the IR
($\ell_{IR}^*=2\pi^2/3$, $\omega_{IR}^*=0$) fixed point solutions. Naturally the Gaussian and the
WF ones has the same behavior as was obtained from direct calculations. However the new IR
fixed point indeed corresponds to the values $\kappa_{IR}^*\to-\infty$ and $\lambda_{IR}^*=0$,
and the linearization in its vicinity gives a negative and a zero eigenvalue, showing that
the fixed point is IR attractive, in accordance with the flows in \fig{fig:phase}.

If one considers more couplings and includes the running of the anomalous dimension one
obtains qualitatively similar results with similar phase space structure as in \fig{fig:phase}
and with the same fixed points. However, in the broken phase the effective potential
has a wide range of flat region in its middle, implying that the Taylor expansion of the
potential as in \eqn{pot} does not converge well. One can obtain more reliable results
if one considers the evolution of the non-expanded potential in \eqn{potev} without a
polynomial ansatz. Nevertheless we continue our investigation by expanding the potential,
similarly to \cite{Liao,Litim_opt,exps_field},
since our goal is to demonstrate the existence of the IR fixed point and to calculate
the corresponding exponents in its vicinity, and not to get better exponents than the
ones obtained by high precision Monte-Carlo calculations \cite{Pelissetto,Hasenbusch}.

\begin{figure}
\begin{center} 
\epsfig{file=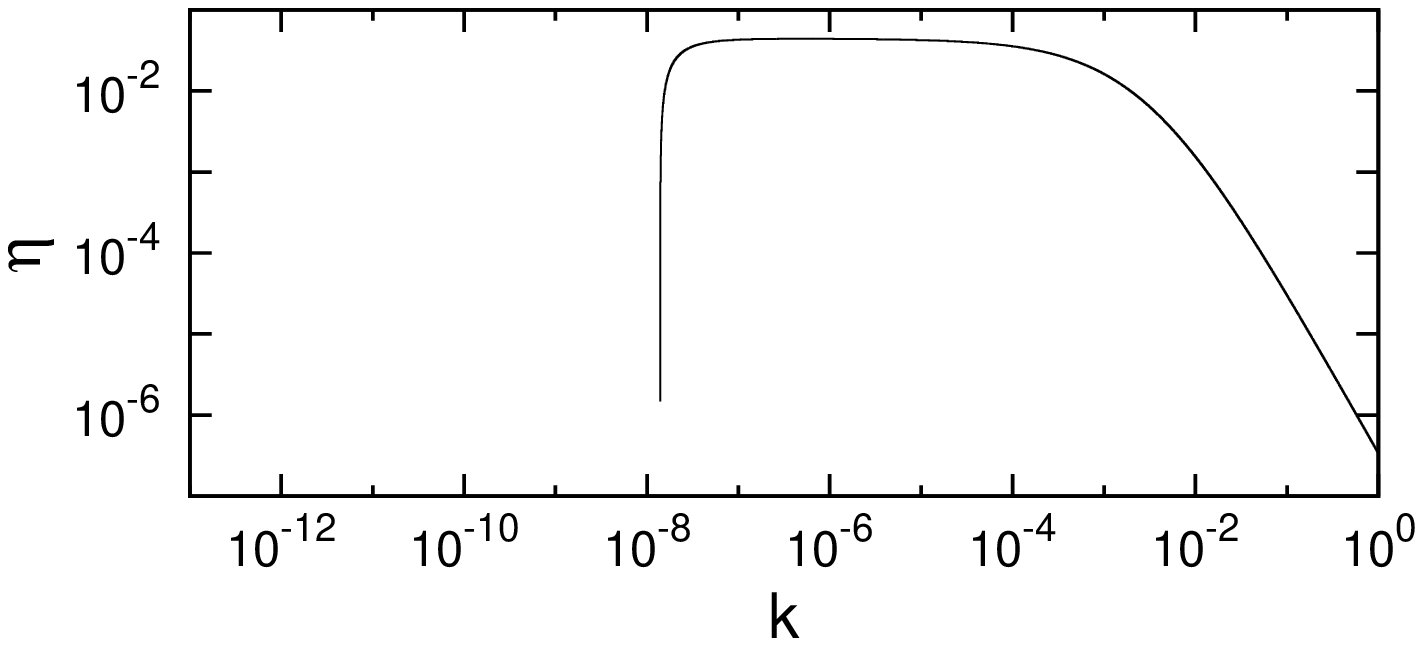,width=4cm,angle=-90}
\epsfig{file=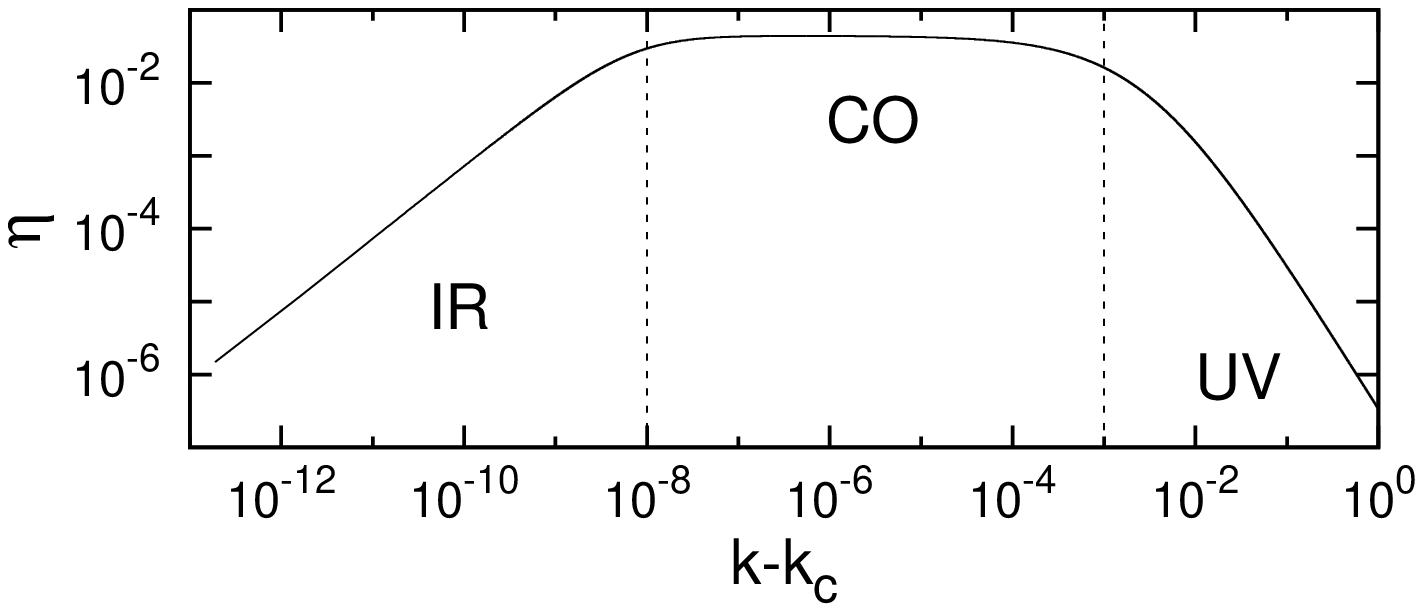,width=4cm,angle=-90}
\caption{\label{fig:eta} The evolution of the anomalous dimension $\eta$ is presented
as the function of the scale $k$ (top) and the shifted scale $k-k_c$ (bottom).
The flow of $\eta$ has a strong singularity as the function of the
scale $k$ at $k=k_c$, but it shows a power law like behavior as the function
of the shifted scale $k-k_c$.} 
\end{center}
\end{figure}

The RG flow for $M$ number of couplings can also drive the evolution in
\eqns{potev}{etak} to degeneracy where $\omega=0$.
The r.h.s. of the RG flow equations become singular, when the degeneracy condition
$\omega=0$ is satisfied at a certain value of $k=k_c$. As $k$ approaches $k_c$ the
anomalous dimension $\eta$ becomes also singular and it tends to zero at $k_c$ abruptly,
as it is shown in the upper figure in \fig{fig:eta}, which
also demonstrates that there are three different scaling regions in its flow.
In order to analyze them we plotted $\eta$ as the function of the shifted scale $k-k_c$ in the
lower figure in \fig{fig:eta}. In the UV region the Gaussian fixed point drives the evolution
of the anomalous dimension.  In this regime it grows according to the power law scaling
$\eta_{UV}\sim (k-k_c)^{-2}$. There is a crossover (CO) scaling region between
$10^{-8}\lesssim k-k_c\lesssim 10^{-4}$ where a plateau appears giving a constant
value for $\eta_{CO}\approx 0.043$ due to the WF fixed point. Going further in the evolution
towards the smaller values of $k$ below $k-k_c\sim 10^{-8}$ one can find a third scaling regime.
It appears a simple singularity in the upper figure, but the shifted scale $k-k_c$ clearly uncovers
the power law scaling of the anomalous dimension there according to $\eta_{IR}\sim(k-k_c)^1$.
This scaling region is induced by the IR fixed point.

The evolution of the other couplings also shows such type of scaling regimes with
similar singularity structure in the IR limit. There the power law
scaling behaviors also takes place as the function of $k-k_c$ with the corresponding exponents.
This shows that the appearing singularities are not artifacts and the RG flows can be
traced down to the value of $k_c$.

We note that one can find such a value of $M$ where the evolution does not stop as
in the upper figure in \fig{fig:eta}, but after the sharp fall $\eta$ continues its RG evolution 
marginally giving a tiny value of $\eta$ there. However, the singular-like fall
possesses the same power-law like behavior as the function of $k-k_c$, for any value of
$M$. It suggests that the value of $\eta$ rapidly falls to zero at $k_c$ and it is due to
the numerical inaccuracy, whether the RG evolution survives the falling and can be traced to
any value of $k$, or the flows stop due to the appearing singularity.
It strongly suggests that the singular behavior with its uncovered IR scaling
for the shifted scale $k-k_c$ is not an artifact but is of physical importance.
We note that one has the same singularity if one solves
the RG equation in \eqn{WRG} without any functional ansatz for the potential \cite{Pangon}.
The RG flows stop at $k_c$ due to the huge amount of soft modes close to the degeneracy. It implies
that during the evolution a dynamical momentum scale is generated, which can be identified
by the characteristic scale of the global condensate in the broken phase.
When the UV value of $\kappa_\Lambda$ is fine tuned to its critical value $\kappa^*_\Lambda$
the dynamical scale $k_c$ tends to zero. The reciprocal of $k_c$ can be identified by
the correlation length $\xi$ of the model.  Such an identification of $\xi$ ensures
that the analysis is performed in the vicinity of the IR fixed point.
In the IR limit $\xi$ diverges similarly to what it does in the presence of any other
fixed point, so the IR physics of the broken phase possesses an IR fixed point.

\begin{figure}
\begin{center}
\epsfig{file=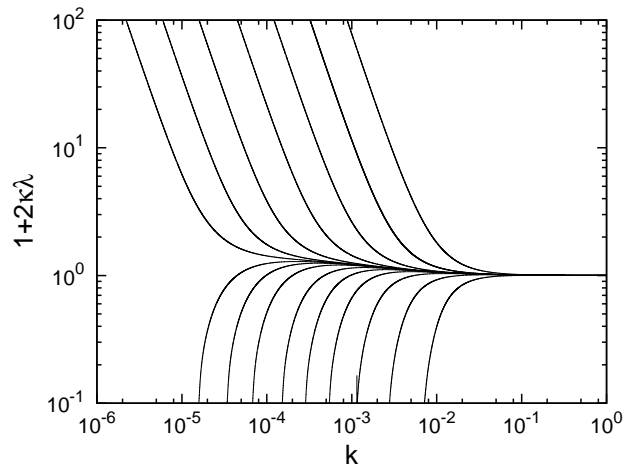,width=6cm,angle=-90}
\caption{\label{fig:corr} The scaling of the degeneracy condition $1+2\kappa\lambda$
as the function of the scale $k$. The diverging curves correspond to the symmetric phase,
while the curves tending sharply to zero correspond to the broken phase.
} 
\end{center}
\end{figure}

In \fig{fig:corr} the quantity $\omega=1+2\kappa\lambda$ is plotted for various initial
conditions.
It shows that in the broken phase, where $\omega\to 0$
the flows break down determining the scale $\xi=1/k_c$ as the size of
the appearing condensate. In the symmetric phase $\omega$
remains finite during the evolution and gives diverging flows in \fig{fig:corr}
according to power-law behavior in the IR. One can identify the reduced temperature $t$
in the $O(N)$ model as the deviation of the UV coupling $\kappa_\Lambda$ to its
UV critical value, i.e. $t\sim \kappa^*_\Lambda-\kappa_\Lambda$.
The correlation length has the power law scaling behavior according to
\beq
\xi\sim t^{-\nu},
\eeq
which characterizes a second order phase transition. The IR defined $\xi$ gives us a
possibility to recalculate the exponent $\nu$ in the vicinity of the IR fixed point.

\section{3\lowercase{d} $O(N)$ model}\label{sec:ond}

\begin{figure}
\begin{center} 
\epsfig{file=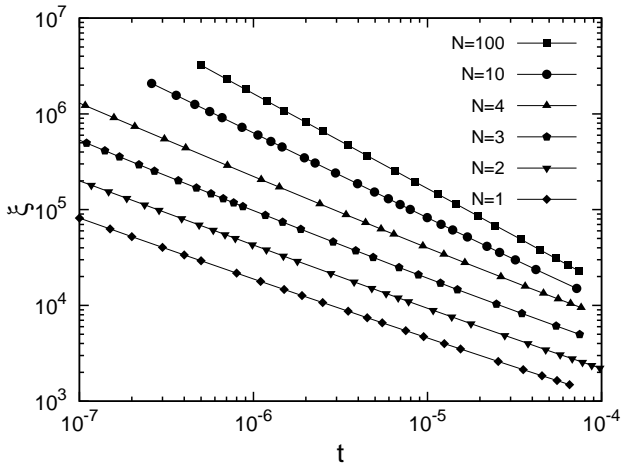,width=6cm,angle=-90}
\caption{\label{fig:d3on} The scaling of the correlation length
$\xi$ as the function of the reduced temperature $t$, for various values of $N$.
} 
\end{center}
\end{figure}

High accuracy calculations exist \cite{Pelissetto,Hasenbusch} for the calculation of the exponent
$\nu$ in the vicinity of the WF fixed point for the 3d $O(N)$ model. Our goal is now to get $\nu$
from the scaling around the IR fixed point. We numerically calculated the scaling of
the couplings and then we determined the critical scale $k_c=1/\xi$ for given UV values
of $\kappa_\Lambda$. The results are plotted in \fig{fig:d3on}.
For a given value of $N$ we obtain power-law behavior for the scaling
of $\xi$, and the slope of the lines provides the exponent $\nu$ in the log-log scale.
The obtained results are listed in \tab{tab:nu}. We denoted the WF (IR) values
of $\nu$ as $\nu_{\mbox{WF}}$ ($\nu_{\mbox{IR}}$), respectively.
\begin{table}
\begin{center}
\begin{tabular}{|c||c|c|c|c|c|c|}
\hline
  $N$ & 1 & 2 & 3 & 4 & 10 & 100 \\
\hline
 $\nu_{\mbox{IR}}$ & 0.624 & 0.666 & 0.715 & 0.760 & 0.883 & 0.990 \\
\hline
 $\nu_{\mbox{WF}}$ & 0.631 & 0.666 & 0.704 & 0.739 & 0.881 & 0.990 \\
\hline
\end{tabular}
\end{center}
\caption{\label{tab:nu} The critical exponent $\nu$ in the $O(N)$ model
for various values of $N$.}
\end{table}
The results show high coincidence. The values $\nu_{\mbox{WF}}$
are taken from results obtained from derivative expansion up to the second
order, since this approximation are the closest to our treatment.
One can conclude that the exponent $\nu$ can be also determined from the
scaling around the IR fixed point, and has the same value as was obtained
around the WF fixed point. This fact might seem quite strange at a first
glance since the two fixed points are well separated in the phase space
according to \fig{fig:phase}. The coincidence may come from the fact that the
condensate of the broken phase has a global feature which accompanies
the whole flow starting from UV to IR. The other exponents do not
necessarily coincide, e.g. anomalous dimension $\eta\approx 0.04$ in the
vicinity of the WF fixed point, while it converges to zero in the IR one.

\section{Dimension-dependence of $\nu$ in the $O(1)$ model}\label{sec:dnu}

As a further test we calculated the exponent $\nu$ for the $O(1)$ model for continuous
dimension. By using $\epsilon$-expansion the dimension dependence of $\nu$ can be
easily obtained \cite{Kleinert,ZJ}. By linearizing the functional RG flow
equations around the WF fixed point the exponent $\nu$ can be calculated for any dimensions
\cite{Ballhausen}. Naturally far from $d=4$ the anomalous dimension grows up, which
requires to take into account the evolution of the wavefunction renormalization.

\begin{figure}
\begin{center} 
\epsfig{file=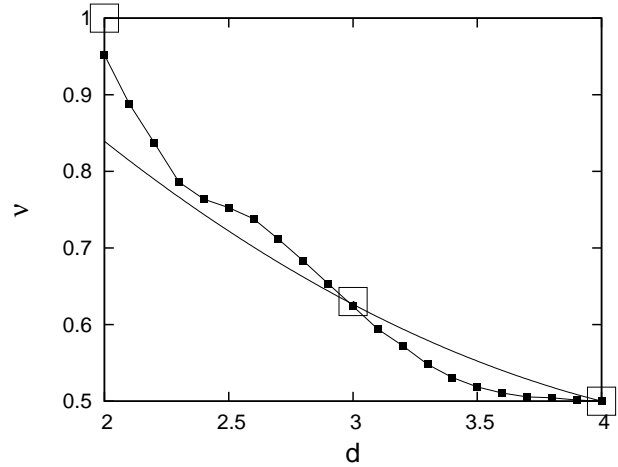,width=6cm,angle=-90}
\caption{\label{fig:dnu} The value of the critical exponent $\nu$ for various
dimension $d$ (filled squares). The continuous curve obtained from $\epsilon$-expansion
\cite{Kleinert}. The empty squares denote the exact values at integer dimensions.
} 
\end{center}
\end{figure}

We determined the exponent $\nu$ by the degeneracy induced scaling around the IR fixed point.
The results are shown in \fig{fig:dnu}.
The exponent $\nu$ calculated around either the WF or the IR fixed points give the same results.
The square at $d=2$ in \fig{fig:dnu} shows $\nu=1$, as the exact results of the
2d Ising model \cite{Gulacsi}. When $d=3$ there are no exact calculations but one
can find a huge amount of articles related to calculating $\nu$ \cite{Pelissetto,Hasenbusch}.
When $d=4$ the mean field calculations give $\nu=0.5$. Our results give better results than
the $\epsilon$-expansion ones for low dimensions, since far from $d=4$ one needs more and more
loop corrections to get reliable results. At around $d\approx 2.5$ there is a wiggle in
\fig{fig:dnu}, which does not appear in \cite{Ballhausen}. This is due to the numerical
inaccuracy in our method coming from the high anomalous dimension in the IR regime
which should be traced to extremely small values of the scale $k$. We expect that by taking into
account higher orders in the derivative expansion or by treating the problem with
full momentum dependence \cite{full_mom} the wiggles could be removed.

These results clearly show that it is possible to determine the power-law scaling of
the correlation length in the vicinity of the IR fixed point. We note that
there is no WF fixed point in case of $d=4$ model. However the degeneracy induced
scaling gives us back the analytic result even in this case, too.

\section{2d $O(2)$ model}\label{sec:o2d2}

The 2-dimensional $O(2)$ model is exceptional among the $O(N)$ models since
it possesses an infinite order KT-type phase transition \cite{KT}. This transition
is characterized by the scaling of the correlation length according to
\beq\label{corrKT}
\log\xi\sim t^{-\nu},
\eeq
with the exponent $\nu=0.5$. The 2d $O(2)$ model belongs to the same universality
class as the 2d XY model \cite{Gulacsi}, the 2d Coulomb gas \cite{cg} the 2d
SG model \cite{sg}, but the equivalent models inherently contain
vortices as elementary excitations which can account for the essential scaling.
The functional RG approach shows KT-type scaling according to \eqn{corrKT} if one
considers the effect of the wavefunction renormalization in the SG model
\cite{Nagy_ZSG}, furthermore it was found that the KT and the IR scaling gives
the same scaling behavior and the same exponent $\nu\approx 0.5$ \cite{Nagy_2010}.
However the 2d $O(2)$ model does not contain any vortices but the scaling
in \eqn{corrKT} can be uncovered \cite{essential,Morris}, although in a rather involved way.

\begin{figure}
\begin{center} 
\epsfig{file=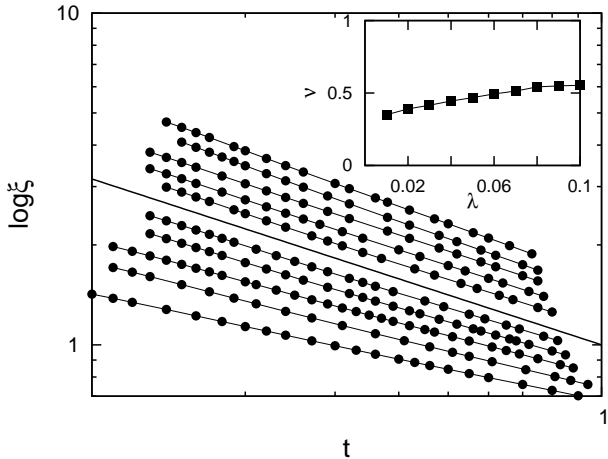,width=6cm,angle=-90}
\caption{\label{fig:o2d2} The KT-type scaling of the correlation length in the
$O(2)$ model in $d=2$. The lines correspond to flows belonging to different
UV values of $\lambda$.
} 
\end{center}
\end{figure}

The 2d $O(2)$ model also contains an IR fixed point, therefore the degeneracy induced
IR scaling also can be used. We plotted the results in \fig{fig:o2d2}.
They show that we can recover the essential scaling according to \eqn{corrKT} in a
very simple way by using our proposed technique, although
the exponent shows a slight dependence on the UV value of $\lambda$ as is depicted
in the inset of \fig{fig:o2d2}. Besides, one can conclude that $\nu\approx 0.5$, in accordance
with other results \cite{Kosterlitz,essential,Nagy_ZSG}.

\section{Summary}\label{sec:sum}

The critical exponent $\nu$ of the correlation length $\xi$ was calculated for the $O(N)$ model
in the IR limit. We showed that there is an IR fixed point in the broken phase
of the model, which induces a degeneracy and stops the RG evolution at a finite momentum scale.
The latter defines a characteristic length scale
which can be identified with the correlation length. This technique of identifying the
correlation length enabled us to determine the exponent $\nu$ around the IR fixed point,
far from the WF one in which vicinity it is usually calculated.
We showed that the value of the exponent $\nu$ agrees well with results
obtained in the vicinity of the WF fixed point for the 3d $O(N)$ model, see \tab{tab:nu}.
The qualitative behavior of $\nu$ for the $O(1)$ model with continuous dimension
presented earlier by $\epsilon$-expansion or linearization around the WF fixed point was
recovered by the IR scaling. Furthermore the essential scaling of the
2d $O(2)$ model with the exponent $\nu\approx 1/2$ was also got back.
The latter model exhibits a KT-type phase transition showing that
this simple technique can be easily applied to describe any types of phase transitions.

The coincidence of the exponent $\nu$ around
the WF and the IR fixed point may come from the fact that the correlation length
in the broken phase characterizes the condensate which occurs there
and consists of a macroscopic population of soft modes showing the global nature
of $\xi$. The anomalous dimension $\eta$ does not agree around the WF and
the IR fixed points. The common feature of the d-dimensional $O(N)$ models for $d>2$
that there is a crossover fixed point, namely the WF fixed point there.
Although the crossover and the IR fixed points
are far from each other in the phase space, we read off the
scaling of $\xi$ from the same trajectories which first approach the crossover
fixed point and then tend towards the IR one.
However the real power of this technique can emerge if there are no crossover scalings.
For example in the 2d $O(2)$ model there is only some remainder of the WF fixed point as
a critical slowing down of the RG flows, or in the 4d $O(1)$ model the WF fixed point
melts into to the Gaussian one. It is shown that the IR scaling can correctly
determine the order of the phase transitions, and the proper value of the exponents $\nu$
in these models, too. A further non-trivial
example is the massive SG model \cite{msg}, and the layered SG model
\cite{Nagy_2010,lsg} where there are no crossover fixed points too, but the IR scaling
uncovers the type of the phase transition and helps us to determine in which
universality class the models belong to \cite{Nagy_per}.

\section*{Acknowledgments}

The author thanks Kornel Sailer for illuminating discussions.
The work is supported by the TAMOP 4.2.1/B-09/1/KONV-2010-0007 and the
TAMOP 4.2.2/B-10/1-2010-0024 projects. The projects are implemented through the New
Hungary Development Plan co-financed by the European Social
Fund, and the European Regional Development Fund.


\begin{thebibliography}{99}
\bibitem{Wetterich}
C. Wetterich, Phys. Lett. B{\bf 301}, 90 (1993).
\bibitem{effRG}
T. R. Morris, Int. J. Mod. Phys. A {\bf 9}, 2411 (1994);
J. Polonyi, Central Eur. J. Phys. {\bf 1}, 1 (2004);
J. Comellas, Nucl. Phys. B {\bf 509}, 662 (1998);
C. Bagnuls, C Bervillier, Phys. Rept. {\bf 348}, 91, (2001);
J. Berges, N. Tetradis, C. Wetterich, Phys. Rept. {\bf 363}, 223 (2002).
\bibitem{Tetradis_1994}
N. Tetradis, C. Wetterich, Nucl. Phys. B {\bf 422}, 541 (1994);
\bibitem{Liao}
S.-B. Liao, J. Polonyi, M. Strickland, Nucl. Phys. B {\bf 567}, 493 (2000);
\bibitem{exps_field}
D. F. Litim, Phys. Rev. D {\bf 64}, 105007 (2001);
L. Canet, B. Delamotte, D. Mouhanna, J. Vidal, Phys. Rev. D {\bf 67}, 065004 (2003);
\bibitem{exps_eps}
R. Guida, J. Zinn-Justin, J. Phys. A {\bf 31}, 8103 (1998);
J. Zinn-Justin, Phys. Rept. {\bf 344}, 159 (2001);
\bibitem{exps_der}
T. R. Morris, Nucl. Phys. B {\bf 495}, 477 (1997);
M. D. Turner, T. R. Morris, Nucl. Phys. B {\bf 509}, 637 (1998);
D. F. Litim, Dario Zappal\'a, Phys. Rev. D {\bf 83}, 085009 (2011).
\bibitem{exps_shoot}
C. Bervillier, J. Phys. Condens. Matter {\bf 17}, S1929 (2005);
\bibitem{Pangon}
V. Pangon, S. Nagy, J. Polonyi, K. Sailer, Int. J. Mod. Phys. A {\bf 26}, 1327 (2011).
\bibitem{exps_ham}
M. Campostrini, A. Pelissetto, P. Rossi, E. Vicari, Phys. Rev. E {\bf 60}, 3526 (1999).
\bibitem{Nagy_2010}
S. Nagy, K. Sailer, arXiv:1012.3007.
\bibitem{Braun}
J. Braun, H. Gies, D. D. Scherer, Phys. Rev. D {\bf 83}, 085012 (2011).
\bibitem{Wetterich_kink}
C. Wetterich, Nucl. Phys. B {\bf 352}, 529 (1991).
\bibitem{Alexandre}
J. Alexandre, V. Branchina, J. Polonyi, Phys. Lett. B {\bf 445}, 153 (1999).
\bibitem{Boyanovsky}
D. Boyanovsky, H.J. de Vega, R. Holman, J. Salgado, Phys. Rev. D {\bf 59}, 125009 (1999);
\bibitem{Litim_opt}
D. F. Litim, Phys. Lett. B {\bf 486}, 92 (2000); Nucl. Phys. B {\bf 631}, 128 (2002);
\bibitem{Litim}
D. F. Litim, JHEP {\bf 0111}, 059 (2001);
D. F. Litim, J. M. Pawlowski, Phys. Rev. D {\bf 66}, 025030, (2002).
\bibitem{full_mom}
F. Benitez, J.-P. Blaizot, H. Chate, B. Delamotte, R. Mendez-Galain, N. Wschebor,
Phys. Rev. E {\bf 80}, 030103 (2009);
F. Benitez, J.-P. Blaizot, H. Chate, B. Delamotte, R. Mendez-Galain, N. Wschebor,
Phys. Rev. E {\bf 85}, 026707 (2012).
\bibitem{Bervillier}
C. Bervillier, A. Juttner, D. F. Litim, Nucl. Phys. B {\bf 783}, 213 (2007);
\bibitem{Tetradis}
N. Tetradis, C. Wetterich, Nucl. Phys. B {\bf 383}, 197 (1992).
\bibitem{Nagy_ZSG}
S. Nagy, I. Nandori, J. Polonyi, K. Sailer, Phys. Rev. Lett. {\bf 102}, 241603 (2009).
\bibitem{Pelissetto}
A. Pelissetto, E. Vicari, Phys. Rept. {\bf 368}, 549 (2002).
\bibitem{Hasenbusch}
M. Hasenbusch, J. Phys. A {\bf 32}, 4851 (1999); Int. J. Mod. Phys. C {\bf 12}, 911 (2001). 
\bibitem{Kleinert}
H. Kleinert, J. Neu, V. Schulte-Frohlinde, K. G. Chetyrkin, S. A. Larin,
Phys. Lett. B {\bf 272}, 39 (1991); Erratum-ibid. B {\bf 319}, 545 (1993).
\bibitem{ZJ}
J. Zinn-Justin, {\it Quantum Field Theory and Critical Phenomena},
(Clarendon Press, Oxford, 1996).
\bibitem{Ballhausen} 
H. Ballhausen, J. Berges, C. Wetterich, Phys. Lett. B {\bf 582}, 144 (2004).
\bibitem{Gulacsi}
Z. Gul\'acsi, M. Gul\'acsi, Advances in Physics {\bf 47}, 1 (1998).
\bibitem{KT}
V. L. Berezinskii, Zh. Eksp. Teor. Fiz. {\bf 61},
1144 (1971) [Sov. Phys.-JETP {\bf 34}, 610 (1972);
J. M. Kosterlitz, D. J. Thouless, J. Phys. C{\bf 6}, 1181 (1973).
\bibitem{cg}
D J Amit, Y Y Goldschmidt and S. Grinstein, J. Phys. A {\bf 13}, 585 (1980);
P. Minnhagen, Rev. Mod. Phys. {\bf 59}, 1001 (1987);
I. Nandori, U. D. Jentschura, K. Sailer, G. Soff, Phys. Rev. D {\bf 69}, 025004 (2004).
\bibitem{sg}
I. N\'andori, J. Polonyi, K. Sailer Phys. Rev. D {\bf 63}, 045022 (2001);
S. Nagy, I. N\'andori, J. Polonyi, K. Sailer, Phys. Lett. B {\bf 647}, 152 (2007);
V. Pangon, Int. J. Mod. Phys. A {\bf 27}, 1250014 (2012);
I. Nandori, arXiv:1108.4643; V. Pangon, arXiv:1111.6425.
\bibitem{essential}
M. Gr\"ater, C. Wetterich, Phys. Rev. Lett. {\bf 75}, 378 (1995);
G. v. Gersdorff, C. Wetterich, Phys. Rev. B {\bf 64}, 054513 (2001).
\bibitem{Morris}
T. R. Morris, Phys. Lett. B {\bf 345}, 139 (1995).
\bibitem{Kosterlitz}
J. M. Kosterlitz, J. Phys C{\bf 7}, 1046 (1974).
\bibitem{msg}
I. Nandori, Phys. Lett. B {\bf 662}, 302 (2008);
I. Nandori, Phys. Rev. D {\bf 84}, 065024 (2011).
\bibitem{lsg}
L. Benfatto, C. Castellani, T. Giamarchi, Phys. Rev. Lett. {\bf 98}, 117008 (2007);
Phys. Rev. Lett. {\bf 99}, 207002 (2007);
E. Babaev, Nucl. Phys. B {\bf 686}, 397 (2004);
J. Smiseth, E. Smorgrav, E. Babaev, A. Sudbo, Phys. Rev. B {\bf 71}, 214509 (2005);
U. D. Jentschura, I. Nandori, J. Zinn-Justin, Annals Phys. {\bf 321}, 1647 (2006);
I. Nandori, J. Phys. A {\bf 39}, 8119 (2006);
J. Kovacs, S. Nagy, I. Nandori, K. Sailer, JHEP {\bf 1101}, 126, 2011.
\bibitem{Nagy_per}
S. Nagy,  Nucl. Phys. B {\bf 864}, 226 (2012).
\end{thebibliography}
\end{document}